\documentclass[a4paper]{jpconf}
\usepackage{graphicx}
\usepackage{amsmath}
\usepackage[T2A]{fontenc}
\usepackage[cp1251]{inputenc}

\begin{document}
\title{On a theory of light scattering from a Bose-Einstein condensate}

\author{V.M. Ezhova, L.V. Gerasimov, and D.V. Kupriyanov}

\address{Department of Theoretical Physics, Peter the Great Saint-Petersburg Polytechnic University, St-Petersburg 195251, Russia}

\ead{EzhovaVic1929@phys-el.spbstu.ru}

\begin{abstract}
We consider a quantum theory of elastic light scattering from a macroscopic atomic sample existing in the Bose-Einstein condensate (BEC) phase. Following the second quantized formalism we introduce a set of coupled and closed diagram equations for the polariton propagator contributing to the $T$ -matrix and scattering amplitude. Our approach allows us to follow important density corrections to the quasi-energy structure caused by static interaction and radiation losses associated with incoherent scattering in the case of near resonance excitation.
\end{abstract}

\section{Introduction}

Light scattering on complex quantum systems is a potentially interesting process for developing various quantum interface protocols organized between the light and matter subsystems. The coherent joint propagation of light and matter wave through a degenerate quantum gas existing in the Bose-Einstein condensate (BEC) phase had been predicted in \cite{Politzer} even before such a matter state had been created in the laboratory. After the first successful experimental realization of BEC in alkali-metal systems in \cite{Cornell,Ketterle}, evident signatures of cooperative dynamics in light scattering on the condensate have been observed in experiments \cite{Pritchard}-\cite{Schneble}. The strong coherent scattering of light on a sample led to the condensate fragmentation; this effect have been explained in \cite{Schneble} in terms of a Kapitza-Dirac diffraction phenomenon. These experiments have encouraged experimental and theoretical efforts towards deeper understanding and realistic description of light scattering process from degenerate bosonic gas \cite{Moor}-\cite{UFN}. In  \cite{Trifonov}-\cite{UFN} a semiclassical approach for light scattering from a BEC considered as an ideal atomic gas have been applied to describe the basic results of experiment \cite{Pritchard,Hilliard}.

In the present report we propose a theoretical approach based on a standard formalism of the quantum scattering theory. We consider BEC in the framework of the Gross-Pitaevskii model \cite{Gross,Pitaevskii} and describe the optical excitation of such a complex quantum system via introducing the self-consistent Green's function analysis.

\section{Scattering problem in conditions of quantum degeneracy }
The quantum-posed description of the photon scattering problem is based on the formalism of the $T$-matrix, which is defined by
\begin{equation}
\label{1}
\hat T(E)=\hat V+\hat V\frac{1}{E-\hat H}\hat V
\end{equation}
where $\hat H$ is the total system Hamiltonian consisting of the nonperturbed part $\hat H_0$ and an interaction term $\hat V$ such that $\hat H=\hat H_0+\hat V$. The energy argument $E$ is an arbitrary complex parameter in Eq.~(\ref{1}). Then the scattering process, evolving from an initial state $|i\rangle$ to the final state $|f\rangle$, is expressed by the following relation between the differential cross section and the transition amplitude, given by the relevant $T$-matrix element considered as a function of the initial energy $E_i$:
\begin{equation}
\label{2}
{\,d\sigma}_{i\to f}=\frac{{\cal V}^2{\omega'}^2}{\hbar^2 c^4{(2\pi)}^2}{|T_{g'\mathbf e'\mathbf k';\,g\mathbf e\mathbf k}(E_i+\rmi\,0)|}^2\,d\Omega
\end{equation}
Here the initial state $|i\rangle=|g;\mathbf{e},\mathbf{k}\rangle$ is specified by the incoming photon’s wave vector $\mathbf k$, frequency $\omega\equiv \omega_k = ck$, and polarization vector $\mathbf e$, and the atomic system populates a particular collective ground state $|g\rangle$. In our case initially $|g\rangle=|\mathrm{BEC}\rangle^N$ performs a collective state of $N$ atoms in the BEC phase described by the Gross-Pitaevskii model.  The final state $|f\rangle=|g';\mathbf{e}',\mathbf{k}'\rangle$ is specified by a similar set of quantum numbers, excepting that $|g'\rangle$ can be now a disturbed condensate state for inelastic channels, and the solid angle $\Omega$ is directed along the wave vector of the outgoing photon $\mathbf k'$. The presence of quantization volume $\cal V$ in this expression is caused by the second quantized structure of the interaction operators. The optical theorem links the total cross section with the diagonal $T$-matrix element
\begin{equation}
\label{3}
\sigma_{\mathrm{tot}}=-\frac{2{\cal V}}{\hbar c}\,\mathrm{Im}\, T_{ii}(E_i+\rmi\,0)
\end{equation}
which gives a convenient tool for the cross-section evaluation via calculation of only one $T$-matrix element for the elastic forward scattering.

In the second quantized representation the interaction term $\hat V$ in Eq.~(\ref{1}), taken in the rotating wave approximation, is given by
\begin{equation}
\hat V=-\sum\limits_n\int\,d^3r\left[\hat{\mathbf d}_{n0}\hat{\Psi}^{\dag}_n(\textbf{r})\hat{\Psi}_0(\textbf{r})\hat{\mathbf E}^{(+)}({\mathbf r})+h.c.\right]%
\label{4}%
\end{equation}
where $\hat{\mathbf d}_{nm}$ is a component of the dipole operator, and where $n=0,\pm 1$ and $m=0$ respectively specify the single atom angular momentum of the exited and non-degenerate ground states for the BEC consisting of atoms with a ${}^{1}S_0$ ground state. $\hat{\mathbf E}^{(+)}({\mathbf r})$ is the positive  frequency component of the electric field operator and the interaction term is described in the long wavelength dipole approximation, see \cite{ChTnDpRcGr}. The operators $\hat{\Psi}_0(\textbf{r})$ and $\hat{\Psi}_n^\dagger(\textbf{r})$ are the second quantized annihilation and creation operators of an atom at position $\mathbf{r}$ respectively in the non-degenerate ground and in a particular excited state and in accordance with the model we take
\begin{equation}
\label{5}
\hat\Psi_0(\textbf{r}){|\mathrm{BEC}\rangle }^N=\Xi(\textbf{r}){|\mathrm{BEC}\rangle }^{N-1}
\end{equation}
where $\Xi(\textbf{r})$ is the ordering parameter ("wavefunction") of the condensate.

The scattering amplitude expressed by "on-shell" $T$-matrix elements contributing to Eqs.~(\ref{2}) and (\ref{3}) and considered for the elastic scattering channel can be expressed by the Green's function of a single particle excitation propagating in the BEC:
\begin{eqnarray}
T_{fi}(E)&=&\frac{2\pi\hbar(\omega_{\mathbf{k}'}\omega_{\mathbf{k}})^{1/2}}{{\cal V}}\iint\,d^3r'\,d^3r\sum\limits_{n',n}{(\mathbf{de'})}^*_{n'0}{(\mathbf{de})}_{0n}\;%
{\mathbf e}^{-\rmi\mathbf k'\mathbf r'+\rmi\mathbf k\mathbf r}\;\Xi^*(\mathbf r')\,\Xi(\mathbf r)%
\nonumber\\
&&\times\left(-\frac{\rmi}{\hbar}\right)\int_0^{\infty}dt\;{\mathbf e}^{\frac{\rmi}{\hbar}(E-E_0^N+\rmi 0)t}\;\rmi\,G_{n'n}(\mathbf{r}',t;\mathbf{r},0)
\label{6}
\end{eqnarray}
where $E_0^N$ is the initial energy of the condensate consisting of $N$ particles and
\begin{equation}
\label{7}
\rmi G_{n'n}(\mathbf{r}',t';\mathbf{r},t)= \langle\mathrm{BEC}|T\Psi_{n'}(\mathbf r';t')\Psi_n^{\dag}(\mathbf r;t)|\mathrm{BEC}\rangle^{N-1}
\end{equation}
is the time ordered (causal) Green's function (propagator) associated with the polariton-type (i.e. superposed between field and atom) quasi-particle excitation propagating through the condensate consisting of $N-1$ particles. The operators contributing to the polariton propagator are the original atomic operators transformed in the Heisenberg representation and dressed by the entire interaction process. We additionally assumed that the condensate itself is a stable system, which cannot be modified by the interaction (\ref{4}) directly without perturbation by an incoming photon.

In accordance with the standard definitions of the diagram technique \cite{BerstLifshPitvsk} the polariton propagator obeys the following
\begin{equation}
\scalebox{0.5}{\includegraphics*{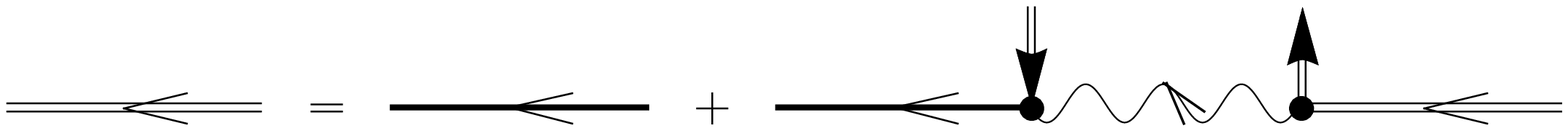}}
\label{8}
\end{equation}
where it is visualized as doubly-straight line dressed by all the interaction processes. The inward and outward vertical arrows image the ordering parameters and form the self-energy part responsible for coherent conversion of the excitation between the field, in which free dynamics is expressed by undressed wavy line, and an atom subsequently recovering to the condensate phase. The degradation of this coherent process is caused by the interaction with the vacuum modes when the excited atom emits a photon spontaneously and escapes coherent dynamics with further drifting through the condensate as a spectator.

The latter process is described by an incomplete atomic propagator (imaged by straight solid line in the diagrams) obeying the following Dyson-type equation
\begin{equation}
\scalebox{0.6}{\includegraphics*{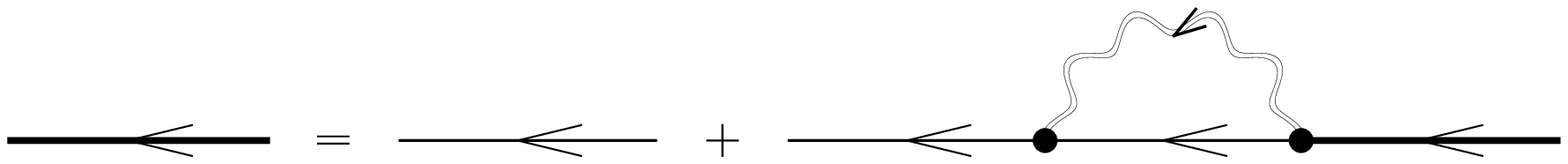}}
\label{9}
\end{equation}
which has to be considered together with the equation for the field propagator
\begin{equation}
\scalebox{0.6}{\includegraphics*{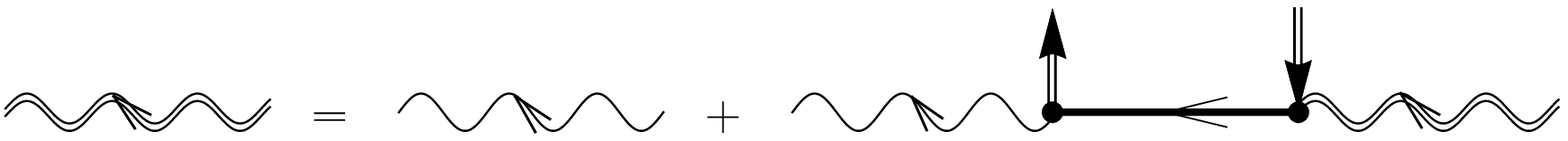}}
\label{10}
\end{equation}
These two diagram equations reproduce the self-consistent dynamics of an atomic dipole interacting with its environment similar the situation in a disordered atomic gas. Any optical excitation created from the condensate has a chance to be incoherently re-emitted into the vacuum modes. As can be verified by precise analytical analysis, both the diagram equations lead to the same structure of polarization operator, dielectric constant and renormalization of the spontaneous decay rate as in disordered atomic gas of the same density. The relevant definitions and calculations have been done in \cite{OurPRA2009}, which we will further apply when solving the basic equation (\ref{8})

\section{Polariton propagator}

For an infinite, homogeneous and isotropic medium, which physically requires that the sample size would be much larger than the radiation wavelength, the solution of Eq.~(\ref{8}) can be found in the reciprocal space as linear combination of the transverse and the longitudinal components with respect to the wave vector argument. In our estimates below, with considering the internal binding energy in the condensate as weak, we will ignore the chemical potential of the condensate as a negligible quantity in comparison with the basic spectral parameters such as spontaneous decay rate, recoil energy etc. Then the Fourier components of the polariton propagator (7) can be expanded as follows
\begin{equation}
G_{nn'}(\mathbf k,\omega)=G_{\parallel}(k,\omega)\,\frac{k_nk_{n'}}{k^2}+G_{\perp}(k,\omega)\left[\delta_{nn'}-\frac{k_nk_{n'}}{k^2}\right]
\label{11}
\end{equation}
where we associated the vector indices in the quasi-particle wave vector $\mathbf{k}$ with the quantum numbers of the atomic excited state.\footnote{The used notation for the momentum argument in the spatial Fourier transform has a signature but should not be directly identified as the wave vector of the scattered photon defined in the previous section.}

Let us first consider and discuss the transverse component of the polariton propagator, which is given by
\begin{equation}
G_{\perp}(k,\omega)=\left[\omega-\omega_0-\frac{\hbar k^2}{2\mathrm{m}}+\frac{4\pi n_0d_0^2}{3\hbar}+\frac{\rmi \sqrt{\epsilon(\omega)}\gamma}{2}-\frac{4\pi n_0d_0^2\omega^2}{\hbar(\omega^2-c^2k^2)}\right]^{-1}
\label{12}
\end{equation}
where "$\mathrm{m}$" is the atomic mass, $n_0=|\Xi(\mathbf{r})|^2=\mathrm{const}$ is the density of atoms in the BEC, $d_0$ is the modulus of the dipole transition matrix element (same for all transitions), and $\omega_0$ is the transition frequency for the atom considered in a two-level approximation, $\gamma$ is its natural radiation decay rate. The dielectric susceptibility $\epsilon(\omega)$ can be analytically calculated for arbitrary atomic density, see \cite{OurPRA2009} for derivation details.

\begin{figure}[t]
\center{\includegraphics[width=1\linewidth]{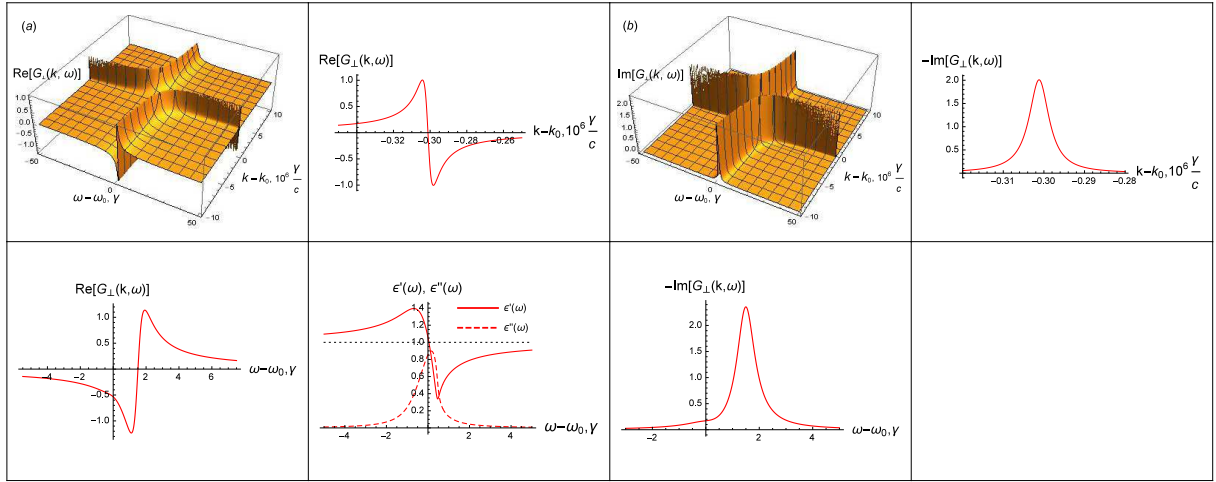}}
\caption{The transverse component of the polariton propagator $G_{\perp}(k,\omega)$ plotted for the density $n_0{(\lambda_0/2\pi)}^3=0.05$, where $\lambda_0=780\, \mathrm{nm}$ is the radiation wavelength at atomic resonance: (a) Real part and (b) Imaginary part. Other graphs show the resolved spectral profiles, taken at the edges of three-dimensional plots (a) and (b), and the dielectric permittivity of the sample $\epsilon(\omega)=\epsilon'(\omega)+\rmi\epsilon''(\omega)$ calculated in \cite{OurPRA2009}.}
\label{fig1}
\end{figure}

In Fig.~\ref{fig1} the transverse component of the polariton propagator Eq.(\ref{9}) is shown as a function of its frequency $\omega$ and wavenumber $k$ for the parameters specified in the caption. As a basic wavelength we use rubidium $D2$-line with $\lambda=780\,nm$.  These plots show that for large frequency detunings the excitation mode approximately approaches the photon-type dispersion law $\omega\sim c|\mathbf{k}|$. The key feature of this result suggests fast propagation of the polariton through the sample and possible strong coherent scattering of light from the sample boundaries where the polariton is created. The coherent scattering could manifest itself in those conditions when a normal atomic gas of the same density would be practically transparent. The strong scattering of light from a BEC of alkali-metal atoms for off-resonant detunings of a few GHz have been observed in experiments \cite{Schneble,Hilliard}.

\begin{figure}[t]
\center\scalebox{0.7}{\includegraphics[width=.8\linewidth]{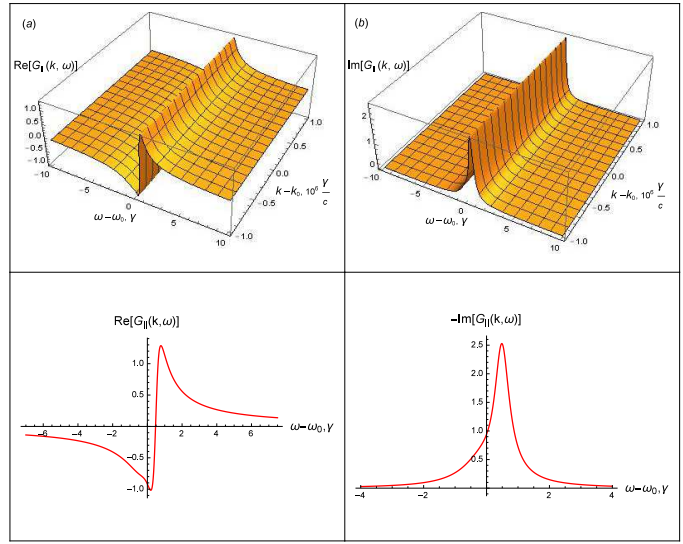}}
\caption{Same as in Fig.~\ref{fig1} but for longitudinal component of the polariton propagator $G_{\parallel}(k,\omega)$.}
\label{fig2}
\end{figure}

With approaching the point of atomic resonance $\omega\to\omega_0$ the optical excitation shows behavior similar to that of a non-condensed disordered atomic gas of the same density. The collective dipole polarization is driven by the propagating field and the environment of proximal dipoles induces the well known red static Lorentz-Lorentz shift  $-4\pi n_0d_0^2/3$ from the atomic resonance. However there is an extra frequency shift, induced by the polarization interaction with the condensate particle, which is given by the last term in the right-hand side of Eq.(\ref{12}). Considering such an exciton-type quasi-particle as motionless with negligible momentum $p=\hbar k\ll \hbar\omega/c$ the dependence on $\omega$ vanishes and this part of the interaction also becomes static. In this limit the transverse component of the polariton propagator coincides with its longitudinal part, such that the excitation process becomes isotropic, and is given by
\begin{equation}
G_{\parallel}(k,\omega)=\left[\omega-\omega_0-\frac{\hbar k^2}{2\mathrm{m}}-\frac{8\pi n_0d_0^2}{3\hbar}+\frac{\rmi\sqrt{\epsilon(\omega)}\gamma}{2}\right]^{-1}
\label{13}
\end{equation}
The spectral behavior of the longitudinal component of the polariton propagator is shown in Fig.~\ref{fig2} for the same parameters as in Fig.~\ref{fig1}.

\section{Conclusion}
In the present paper we have shown that light propagates through a degenerate atomic gas existing in the BEC phase as a superposed state of field and matter - polariton wave. The polariton, created on the sample boundaries, can lead to strong coherent scattering of light from the sample edges in a broad frequency spectrum. This potentially valuable property could be applied for developing optomechanical quantum interface schemes between the non-classical light and condensate with using superfluidity as a main macroscopic quantum property of such a matter state.

\section*{Acknowledgements}

This work was supported by the RFBR grant 15-02-01060, by the Foundation "Dynasty" and by the External Fellowship Program of RQC.

\renewcommand{\refname}{\vspace{-1cm}}

\end{document}